# Visualization of variations in human brain morphology using differentiating reflection functions

**Gibby Koldenhof, May 2003**

**Institute for Computer, Communication and Media Technology (ICIM)**
G.Koldenhof@icim.org (1)

**University Medical Center Utrecht (UMCU)**
G.Koldenhof@azu.nl (2)

**Abstract**

Conventional visualization media such as MRI prints and computer screens are inherently two dimensional, making them incapable of displaying true 3D volume data sets. By applying only transparency or intensity projection, and ignoring light-matter interaction, results will likely fail to give optimal results. Little research has been done on using reflectance functions to visually separate the various segments of a MRI volume. We will explore if applying specific reflectance functions to individual anatomical structures can help in building an intuitive 2D image from a 3D dataset. We will test our hypothesis by visualizing a statistical analysis of the genetic influences on variations in human brain morphology because it inherently contains complex and many different types of data making it a good candidate for our approach.

**Keywords:** MRI, Volume rendering, medical visualization, BRDF, specular reflection overlap

## Introduction

Conventional visualization media such as Magnetic Resonance Imaging (MRI) prints and computer screens are inherently two dimensional, making them incapable of displaying true 3D volume data sets. Attempts to overcome this by imposing transparency or intensity projection [1] are likely to be insufficient because no light-matter interaction is integrated. Previous research has shown that a more intuitive visualization can be attained by using color [32], emittance [33], shading parameters and texturing [34], index of refraction [35], by separating spectral information for objects [2,3] or by combining volume rendering algorithms with surface rendering algorithms [4,5,6]. Another interesting approach is to use Non Photo-realistic Rendering (NPR) [48, 49, 50]. Little research however has been done on using reflectance functions [40] (also see [7-12]) to visually separate the various segments of a MRI volume. *Surface graphics* has received a lot of attention in both the academic and commercial world in the last decade while *volume graphics* is still a relative newcomer in the computer graphics field. The inherent problem of volume graphics is that visualization of inner structures of objects requires an alternative rendering approach as opposed to surface visualization. While the surface based rendering implementations have focused on powerful global illumination schemes (ray tracing, radiosity, photon mapping [57], etc, see [15] for an overview) and advanced local illumination algorithms (Ward [41], Neumann et al [12], Ashikhmin [11], see also [20-22]) the volume graphics community has mainly focused on increasing speed of visualization [17,18], transfer functions [51,52], interpolation schemes [13,14] and gradient and normal estimation schemes [23-31].

The idea behind separating reflection models and mixing them for different objects stems from two principles; Humans are able to quickly distinguish, and therefore localize, objects through recognizing their reflectance properties (psychophysically based reflection) [36-38], the assumption being that if we link each object to a different reflection model inter-object relationships should become clearer. The second principle is a simple mathematical one: given two overlapping surfaces with the same orientation and tangent, varying reflection models will not overlap to the same degree in the specular reflection lobe as two reflection models of one kind, thus avoiding the surfaces to 'blur' into one.

Psychophysically based reflection modeling and analysis is still a relatively young research area and little study has been done in this field. Psychophysically based reflection modeling is a perceptually based science and recognizes that the human visual system is able to differentiate, among other, a number of variations in surface reflections. Hunter [36] observed that there are at least six different visual phenomena related to specular reflection. In Hunter's terminology, the separation of elements is based on two perceptually driven factors:

1. Specular gloss (the perceived brightness associated with the specular reflection from a surface)
2. Contrast gloss (the perceived relative brightness of specularly and diffusely reflecting areas).

Fleming et al [37] noted that under normal viewing conditions, humans find it easy to distinguish between objects made out of different materials such as plastic, metal, or paper. Non-textured materials such as these have different surface reflectance properties, including lightness and gloss. The regularity of real-world illumination leads to informative relationships between the reflectance of a surface and certain statistics of an image of that surface. Dror [38] studied these relationships empirically and analytically using a parameterized Ward [41] reflectance model. Building on the foundation of these studies, we will investigate if differentiating reflection models, i.e. applying a different reflection model to each object, will indeed help an observer to more quickly identify a specific object within a structure of multiple objects in the context of MRI data. We postulate that if, in a collection of overlapping objects, different reflection models are applied to each object, humans will find it easier to visually separate one object from the other. We will examine to what extent reflection models should differ and will introduce a measure that can help in selecting combinations of models, as well as examine if using one single reflection model for all objects is indeed significantly worse in terms of object visibility. Fundamentally, we ask ourselves: *can differentiating reflectance functions per anatomical object help a neurological expert to more quickly localize regions and visually separate anatomical features, in volumetric data?*

This paper is organized as follows; first, we will explore the mathematical background of reflection and reflection models. After selecting a number of reflection models, an experiment is conducted using synthetic data in order to reduce the number of candidates. The candidates are then applied to a real world situation by visualizing a statistical analysis of the genetic influences on variations in human brain morphology for a further qualitative experiment.

## Nomenclature

| Definition | Symbol |
| --- | --- |
| Wavelength | $\lambda$ |
| Flux | $\Phi$ |
| Irradiance | $L_i$ |
| Radiance | $L_o$ |
| BRDF (specular) | $f_s$ |
| BRDF | $f_\lambda$ |

## Reflectance theory

Color and specular reflection are two fundamental elements for recognizing materials. Color is related to a surface's *spectral* reflectance properties while specular reflection is a function of a surface's *directional* reflectance properties. The *perception* of surface properties can be categorized into geometric cues (shape) and photometric cues (optics). Since the geometry is based on medical data, and therefore static, we will only investigate photometric variations. Photometric cues can be divided into spatially varying spectral reflectance (texture) or spatially uniform reflectance; the uniform reflectance can further be subdivided into absorption, transmission (refraction and scattering) and reflection. Reflection is the sum of diffuse, specular and glossy reflection elements. Diffuse reflection is non-directional (or rather *all-directional*), specular reflection is strongly directional and glossy reflection exhibits moderate directional scattering; more light is reflected in a restricted part of the hemisphere, usually around the perfectly specular scattering direction. Since diffuse reflection is non-directional and only *perceived* as brightness [37] it cannot help in identifying structures. Only the specular and glossy parts are important for identifying surface characteristics, since only specular reflections are directional we will solely focus on the latter.

Reflections are mathematically defined with a Bi-directional Reflectance Distribution Function (BRDF) [19]. Evaluating the BRDF will yield the ratio of reflected light on a local surface given the incident and scatter directions, it is therefore a *local illumination* model. Determining the scatter direction, given the incident direction, a Probability Distribution Function (PDF) is employed that expresses the probability of flow direction of light in a scene. The PDF is part of a larger framework referred to as *global illumination*, or the more general *light transport system*. To properly model light interaction in a scene, all possible light paths need to be sampled yielding either Monte Carlo integration, (bi-directional) path tracing, photon mapping or other sampling strategies (see [40] for an excellent introduction to sampling strategies). Since we are only interested in measuring specular light paths, we employ the ray-tracing [47] algorithm as our light transport system; it can fully sample specular reflection light paths and is the most computationally cost effective solution for specular sampling.

---

(1) Institute for Computer, Communication and Media Technology, Berkenweg 11, 3818 LA Amersfoort, The Netherlands, room 1020, phone +31 334228908

(2) Rudolf Magnus Ins. of Neuroscience, University Medical Center, Room A01.126, Heidelberglaan 100, 3584 CX Utrecht, The Netherlands, phone +31-302507121

**BRDF definition**

The BRDF of a surface defines its reflectance by specifying what proportion of the light incident from each possible illumination direction is reflected in each possible observation or view direction [40]. The BRDF (see Figure 1) is a function of two directions expressed in spherical coordinates, wavelength and the bi-normal and tangent position denoted by *u* and *v* parameterized in surface space. Traditionally it is denoted by:

$$f_\lambda(u,v,\theta_i,\phi_i,\theta_s,\phi_s) \quad (1.1)$$

In this paper, we will use a more compact formulation as our basic notation:

$$f_\lambda(x,\Theta_i \rightarrow \Theta_s) \quad (1.2)$$

Where $\lambda$ is used to indicate that the BRDF depends on wavelength, the parameter $\Theta_i$ represents the incoming light (incident) direction in Cartesian coordinates, $\Theta_s$ represents the scatter (outgoing) direction in Cartesian coordinates and *x* represents the surface position parameterized in local surface space.

We use the arrow ( $\rightarrow$ ) notation to indicate light or energy flow direction (in this case from incident to scatter).

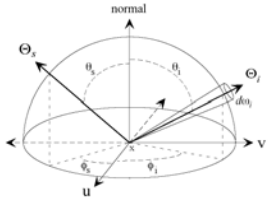

**Figure 1, BRDF definition**

For a BRDF to be classified as *physically plausible* [39], two constraints should be met. First, the BRDF must obey *energy conservation*; for any illumination, total reflected energy must be less than or equal to total incident energy. Second, the BRDF must satisfy the *Helmholtz reciprocity theorem* [66], which guarantees symmetry between incident and scatter directions. Reciprocity requires that:

$$f_\lambda(x,\Theta_i \rightarrow \Theta_s) = f_\lambda(x,\Theta_i \leftarrow \Theta_s) \quad (1.3)$$

Reciprocity is important because it allows for the backward tracing of light as happens in ray-tracing algorithms. Suppose that the surface is illuminated by a beam from direction $\Theta_i$, energy conservation means that the *albedo*, that is the fraction of the total reflected power cannot be greater than one:

$$a_\lambda(\Theta_i) = \int_\Omega f_\lambda(\Theta_i \rightarrow \Theta_s)\cos\theta_s d\omega_s \theta_s \leq 1 \quad (1.4)$$

Where $\Omega$ is the incoming hemisphere and $d\omega$ the (differential) solid angle, the solid angle is defined as:

$$d\omega = \frac{dA}{r^2} \quad (1.5)$$

Given that *r* is the radius of the unit sphere and *dA* the area on that sphere.

*Anisotropy* refers to BRDF models that describe reflectance properties that exhibit change with respect to rotation of the surface about the surface normal vector as opposed to *isotropy* where there is no change in reflectance with respect to normal vector rotation.

Traditionally a BRDF includes a diffuse and a specular component, for our purposes we will only focus on the specular component of each BRDF. Please note we have omitted all non-specular component computations in all BRDF models mentioned in this paper. Also, note that we will drop the wavelength from now on in our equations for simplicity and neglect polarization. Our BRDF notation therefore becomes $f_s(x,\Theta_i \rightarrow \Theta_s)$ for the remainder of this paper.

**BRDF and Fresnel behavior**

Although not included in the definition of "physically plausible" *Fresnel behavior* [40] is important if striving for accurate light-matter interaction. The Fresnel equation predicts a color shift of the specular component at grazing angles. The Fresnel effect is wavelength dependent; it explains the variation in colors seen in specular regions - particularly on metals. It also explains why most surfaces approximate mirror reflectors when the light strikes them at a grazing angle. Although Fresnel behavior is not part of the BRDF *definition*, it should be included in the BRDF *implementation* if striving for more physically plausible interaction. Fresnel reflection is formulated as:

$$F(\Theta_i) = \frac{1(g-c)^2}{2(g+c)^2}\left(1 + \frac{[c(g+c)-1]^2}{[c(g-c)+1]^2}\right) \quad (1.6)$$

With $c = (\Theta_s \bullet \frac{\Theta_i + \Theta_s}{|\Theta_i + \Theta_s|})$ and $g = \sqrt{n^2 + c^2 + 1}$ where *n* is the index of refraction.

Most BRDF implementations use an approximation introduced by Schlick [53] that has less than one percent error compared to equation (1.6), it is formulated as:

$$F(\Theta_i) \approx F_o + (1-F_o)(1-\cos\theta_i)^5 \quad (1.7)$$

Where $F_o$ is the value of the real (of the complex) Fresnel coefficient at normal incidence.

**BRDF and radiometry**

The BRDF for a surface point *x* is defined as the ratio of outgoing radiance $L_o$ and differential irradiance $L_i$ for an incident direction, adjusted by the cosine of the incident polar angle $\theta_i$ and the solid angle $d\omega$ of the incident.

$$f_s(x,\Theta_i \rightarrow \Theta_s) = \frac{L_o(x \rightarrow \Theta_s)}{L_i(x \leftarrow \Theta_i)\cos\theta_i d\omega} \quad (1.8)$$

The BRDF has units of inverse steradian ( $sr^{-1}$ ). It can vary from zero (no light reflected) to infinity (unit radiance in an exit direction).

The radiance leaving a surface due to irradiance in a particular direction is obtained from the definition of the BRDF:

$$L_o(x \rightarrow \Theta_s) = f_s(x,\Theta_i \rightarrow \Theta_s)L_i(x \leftarrow \Theta_i)\cos\theta_i d\omega \quad (1.9)$$

We are however interested in the radiance leaving a surface irrelevant of direction of irradiance, this is obtained by integrating over contributions from all incoming directions:

$$L_o(x \rightarrow \Theta_s) = \int_\Omega f_s(x,\Theta_i \rightarrow \Theta_s)L_i(x \leftarrow \Theta_i)\cos\theta_i d\omega \quad (1.10)$$

A light source is specified by its geometry and an emission distribution function (EDF). The EDF defines the *self-emitted radiance*, $L_e(x \rightarrow \Theta)$, emitted from a point *x* on a light source into direction $\Theta$ .

The *emittance* or self-emitted radiosity is defined as the self-emitted radiance integrated over the hemisphere:

$$B_e(x) = \int_\Omega L_e(x \rightarrow \Theta)\cos\theta d\omega \quad (1.11)$$

The self-emitted *flux* for a light source '*light*' with surface area $A_l$ is given by:

$$\Phi_e^{light} = \int_{A_l} B_e(x)dA \quad (1.12)$$

Combining all equations yields the *rendering equation* [8] which expresses the radiance leaving a surface in a point *x* towards the direction $\Theta_s$ :

$$L_o(x \rightarrow \Theta_s) = L_e(x \rightarrow \Theta_s) + \int_\Omega L_i(x \leftarrow \Theta_i)f_s(x,\Theta_i \rightarrow \Theta_s)\cos\theta_i d\omega \quad (1.13)$$

The unknown irradiance $L_i(x \leftarrow \Theta_i)$ in equation (1.13) originates from outgoing radiance on other surfaces, revealing the recursive nature of the integral equation. In vacuum, i.e. without participating media, radiance leaving a surface will travel unaffected until it reaches another surface. The relation between the irradiance $L_i$ in *x* and the outgoing radiance $L_o$ is given by the *raycast* operation:

$$L_i(x \leftarrow \Theta_i) = L(y \rightarrow \Theta_i) \quad (1.14)$$

The raycast operation finds the nearest intersection *y* for a ray shot from *x* in the direction $\Theta_i$ :

$$y = \text{raycast}(x \rightarrow \Theta_i) \quad (1.15)$$

The rendering equation (1.13) is expressed in terms of incoming and outgoing directions with respect to a surface point *x*, rewriting it using surface points only we obtain (see Figure 2):

$$L_o(x \rightarrow z) = L_e(x \rightarrow z) + \int_\Omega L_i(y \rightarrow x)f_s(y \rightarrow x \rightarrow z)G(x \leftrightarrow y)dA_y \quad (1.16)$$

Where the geometry term, $G(x \leftrightarrow y)$, is defined in local frame space as:

$$G(x \leftrightarrow y) = V(x \leftrightarrow y) \frac{\cos(\theta_i)\cos(\theta_r)}{\|x-y\|^2} \quad (1.17)$$

Where $V(x \leftrightarrow y)$ is the visibility between point $x$ and $y$ (0 if invisible, 1 if visible). Note that the integration over the projected solid angle is replaced by an integration over the surfaces in the scene.

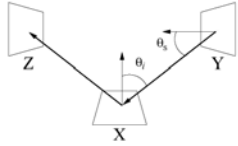

Figure 2, rendering equation using surface points

## BRDF model selection

For our purposes, we are looking for BRDF models that will capture most of the following characteristics:

o  **Physically plausible**: as defined by Lewis, this refers to the model obeying energy conservation and reciprocity. Physically plausible models will behave in ways that are more accurate and will likely help the viewer recognize the material through enhanced realism.

o  **Fresnel behavior**: specularity should increase as the incident angle goes down. Fresnel behavior is important for accurate and realistic light-matter interaction.

o  **Anisotropic**: anisotropy refers to BRDF models that describe reflectance properties that exhibit change with respect to rotation of the surface about the surface normal vector -anisotropic behavior will most probably yield small specular lobe overlapping.

o  **Popular:** BRDF models that are widely accepted and seen in many implementations.

We selected the following nine BRDF models:

**(1)** Phong [55], although not anisotropic or physically plausible, it is without doubt the most popular BRDF. **(2)** Strauss [56], neither anisotropic nor physically plausible, it is an empirically derived substitute for Phong[1]. **(3)** Schlick-Lewis [53], the Schlick BRDF is an attempt to define a BRDF that is qualitatively similar to the Phong BRDF but is faster, adding Lewis' modification allows it to conserve energy. **(4)** Ward [41], Ward designed a model that is quick to evaluate, physically plausible and supports isotropic and anisotropic reflections through an elliptical gaussian model.[2] **(5)** Cook-Torrance [44], the Cook-Torrance-Sparrow BRDF although only isotropic is a complex yet elegant BRDF that is physically plausible (physically based). The model uses a Beckmann micro-facet (geometric attenuation) distribution function, Blinn's [16] geometric shadowing term and Fresnel reflections. **(6)** Poulin-Fournier [46], Poulin and Fournier also make use of a geometric attenuation factor as used in the Cook-Torrance model. The model was developed as a means to achieve a specific appearance and in that sense it has a few deficiencies - It is however not physically plausible and neither does it allow for anisotropic reflections.[3] **(7)** He-Torrance [45], the He-Torrance BRDF offers a comprehensive physical model that addresses deficiencies found in earlier models such as polarization and directional Fresnel effects. It also includes a more detailed formulation of a statistically-described surface geometry: roughness computation that is dependent upon the incident and reflected angles, and a geometric attenuation factor with better continuity than the one used by Cook and Torrance. Yet, it does not allow anisotropic behavior. **(8)** Lafortune [42], the Lafortune generalized cosine lobe analytical BRDF model, although not physically plausible is quite convincing and includes anisotropic behavior[4]. **(9)** Ashikhmin [11], Ashikhmin designed a simple Phong like model with intuitive parameters that is quick to evaluate, physically plausible and supports isotropic and anisotropic reflections. Fresnel behavior is modeled through Schlick's approximation.

An overview of scores based on our criteria for each BRDF model sorted on score is given in Table 1.

| # | BRDF | Properties | Score |
|---|---|---|---|
| 6 | Poulin-Fournier | None | 0 |
| 1 | Phong | Popular | 1 |
| 8 | Lafortune | Physically plausible | 1 |
| 2 | Strauss | Popular | 1 |
| 3 | Schlick-Lewis | Popular, Physically plausible | 2 |
| 5 | Cook Torrance | Physically plausible, Fresnel behavior | 2 |
| 7 | He-Torrance | Physically plausible, Fresnel behavior | 2 |
| 9 | Ashikhmin | Physically plausible, Fresnel behavior, anisotropic | 3 |
| 4 | Ward | Physically plausible, Fresnel behavior, anisotropic, Popular | 4 |

Table 1, BRDF scores

---

[1] The Strauss model simulates Fresnel behavior
[2] The original equation is implemented, not the approximation suggested in Ward's paper.
[3] The specular term is integrated numerically using Simpson's rule instead expanding in terms of Chebyshev polynomials.
[4] Only a one-lobe model was implemented

To get an impression of the specular reflection distinctiveness of each individual BRDF we rendered a MRI dataset of a west-European 30-year-old Caucasian male. The dataset was obtained from a T1-weighted scan (3D fast field echo scans with 160-180 1.2 mm contiguous coronal slices) of the whole head, made on a Philips NT 1.5 Tesla Gyroscan scanner. The MRI dataset was rendered against a random colored background to avoid the Gelb [67] effect, i.e. the inclination for an isolated surface to appear lighter in color than it really is (more formally researched by Wallach [68]). Blur was added to the background to separate the dataset from the background in depth. All parameters of the non-physically plausible BRDF models were modified to reflect the physically plausible BRDF models as much to ensure fair comparison. See appendix B for parameter settings. A Lambertian diffuse constant of 0.2 was added for improved geometric contour.

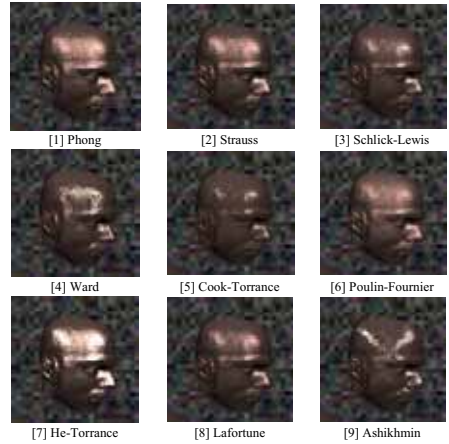

[1] Phong        [2] Strauss        [3] Schlick-Lewis
[4] Ward         [5] Cook-Torrance  [6] Poulin-Fournier
[7] He-Torrance  [8] Lafortune      [9] Ashikhmin

Figure 3, BRDF impression overview

## BRDF flux overlap

To quantify specular overlap an empirical measurement is required to measure flux overlap between two BRDF models in the specular lobe region. The premise being that if overlap is complete, the underlying reflection lobe is invisible; hence, we require a value that indicates how well two BRDF models would perform if combined in one rendering given that lobes should have minimum overlap.

### Flux overlap definition

The rendering equation (1.16) provides an expression for the radiance leaving a single point on a surface. A *measurement* is defined as the response of a certain sensor that combines a set of radiance values. We are interested in measuring specular radiant *flux* through a pixel in an image. Note that we define *pixel* loosely, strictly speaking we are measuring *flux in terms of power*. We define our measurement by a response function $W_e(x \to \Theta)$ that defines the sensitivity of a radiance sensor placed in the image lattice for an area $A$. The total response of the sensor gives the measurement:

$$\Phi_e^{pixel} = \int_A \int_\Omega W_e(x \to \Theta) L_i(x \leftarrow \Theta) \cos\theta \, d\omega \, dA \quad (1.18)$$

Note that the total response is defined as an outgoing quantity, and that it is combined with the incoming radiance function.

The response function itself is formulated as:

$$W_e(x \to \Theta) = \begin{cases} \delta(x - camera) & \text{if } y \in A_{pixel} \\ 0 & \text{otherwise} \end{cases} \quad (1.19)$$

Where *camera* is the viewpoint, $\delta$ the Dirac impulse function and $A_{pixel}$ the image lattice.

The response function corresponds to a *pinhole camera,* which assumes that the aperture of the camera is a single point. The response function is given in terms of two points in the scene, $x$ and $y$; it differs from zero only when $x$ is equal to the camera, and $y$ is located on the image plane. The response function measures the radiant flux through a pixel[5].

Rewriting the measurement equation (1.18) using surface points by transforming the solid angle into an area and including the geometry term (equation (1.17)) we obtain:

$$\Phi_e^{pixel} = \int_{A_x, A_y} W_e(x \to y) L_o(x \leftarrow y) G(x \leftrightarrow y) \, dA_x \, dA_y \quad (1.20)$$

---

[5] The human eye responds to radiance rather than to flux, pixel flux is usually converted through a linear operator to a normalized radiance value.

A *flux overlap* is defined by a commutative function, $f_{FO}$ which weighs flux intensity pixels percentage wise in all areas that do not overlap per reflection lobe and adds a summation of the Euclidian distance between flux intensities in the areas that *do* overlap. Measuring the final flux overlap between two BRDF models is done using a standard mean square error metric (MSE) for the complete rectangular lattice with a commutative total flux overlap function, $f_{TFO}$. A non-overlapping area is defined as a region in which flux values are higher than a certain threshold, essentially creating a binary mask of the reflection lobe. To the same extent, an overlapping area is where both flux values are higher than the latter threshold. The total flux overlap function is formulated as:

$$f_{TFO}(A,B) = \frac{1}{mn}\sum_{y=0}^{m-1}\sum_{x=0}^{n-1}[f_{FO}(a_{xy},b_{xy})]^2 \qquad (1.21)$$

Where *A* is the primary flux lattice, *B* the secondary lattice, *a* the first measured flux value and *b* the second and *m* and *n* are the dimensions of the lattice. Note that $a \in A$ and $b \in B$. The flux overlap function is formulated as:

$$f_{FO}(a,b) = \begin{cases} \dfrac{c}{\text{area}(A)} & \text{if } \dfrac{c}{\text{area}(A)} \leq \dfrac{d}{\text{area}(B)} \\ \dfrac{d}{\text{area}(B)} & \text{if } \dfrac{c}{\text{area}(A)} > \dfrac{d}{\text{area}(B)} \end{cases} \qquad (1.22)$$

Where the total active flux area is defined by the mentioned threshold *t*, and computed as:

$$\text{area}(V) = \sum_{y=0}^{m-1}\sum_{x=0}^{n-1} R_{xy}(v) \text{ with } R(v) = \begin{cases} 1 & \text{if } \Phi_e^v < t \\ 0 & \text{otherwise} \end{cases} \qquad (1.23)$$

Individual flux intensities are computed by:

$$c = \begin{cases} \Phi_e^a & \text{if } \Phi_e^a \notin B \\ |\Phi_e^a - \Phi_e^b| & \text{if } \Phi_e^a \in B \text{ and if } \Phi_e^a > \Phi_e^b \\ 0 & \text{otherwise} \end{cases} \qquad (1.24)$$

$$d = \begin{cases} \Phi_e^b & \text{if } \Phi_e^b \notin A \\ |\Phi_e^a - \Phi_e^b| & \text{if } \Phi_e^b \in A \text{ and if } \Phi_e^a < \Phi_e^b \\ 0 & \text{otherwise} \end{cases} \qquad (1.25)$$

Where $\Phi_e^a$ and $\Phi_e^b$ are measured using equation (1.20) for lattice *A* and *B* respectively, and the Euclidian distance between two flux intensities is measured by $|\Phi_e^a - \Phi_e^b| = \sqrt{[\Phi_e^a - \Phi_e^b]^2}$. See Figure 5 for pseudo code of equations (1.22) through (1.25).

Note that we cannot simply use Euclidian distance measurement between two lobes in all regions; if the height, expressed in flux, never reaches the secondary lobe it will be invisible, hence we cannot sum the distance for that particular lobe (see Figure 4A). If the secondary lobe extends the primary lobe, the flux overlap equation will return the height difference between primary and secondary (see Figure 4B). Regions that do not overlap receive the flux amount for each position on the lattice, e.g., if two lobes do not overlap to a certain extent, both will receive higher scores (see Figure 4C). The lowest of the two scores is returned since the area that does not overlap will determine the actual visible delta between lobes.

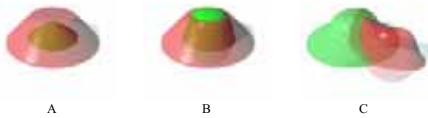

A         B         C

**Figure 4, variations in lobe overlap**

**Flux overlap test**

Given our selection of BRDF models and given the total flux overlap equation we can measure how well two BRDF models fit together. A *good fit* is obtained if the specular overlap is as little as possible, i.e. given a 2D isotropic lattice in Cartesian space, our goal is to find a flux overlap value that is as small as possible, indicating the least possible flux overflow between two different BRDF models in turn indicating the best combination of BRDF models.

We conducted an analysis (see Figure 6 for pseudo code) on all possible permutations of specular lobes in the BRDF models we selected in the previous chapter. Each BRDF was evaluated in combination with another BRDF and the error was measured in terms of difference in total flux overlap error (equation (1.21)). Note that low error values correspond to more overlap; hence, we are looking for high error values.

Each lattice was highly super sampled to avoid possible aliasing errors and a threshold (t) of 0.0001 (equation (1.23)) was used to remove any possible noisy flux values. Tests were performed with three EDF sources (lights) and various geometric primitives to assure all possible tangent orientations of surfaces were included. All geometric primitives were defined as implicit functions allowing for a high degree of surface differentation. Samples were taken at three color-bands representing spectral samples at 700.0 nm, 546.1 nm and 435.8 nm Tests were performed on a 666Mhz Pentium-3 computer running Linux, taking approximately 20 hours.

```
overlap1 = 0, overlap2 = 0, area_A = 0, area_B = 0

for each x in N
  for each y in M
    flux1 = lattice_A[x][y]
    flux2 = lattice_B[x][y]

    distance = sqrt(flux1 * flux1 - flux2 * flux2);

    if flux1 > threshold
      area_A += 1

    if flux2 > threshold
      area_B += 1

    if flux1 > threshold
      if flux2 > threshold
        if flux1 > flux2
          overlap1 += distance
        else
          if flux2 > flux1
            overlap2 += distance
      else
        overlap1 += flux1
    else
      if flux2 > threshold
        overlap2 += flux2
  end for
end for

overlap1 = overlap1 / area_A
overlap2 = overlap1 / area_A

if overlap1 > overlap2
  error = overlap2
else
  error = overlap1
```

**Figure 5, flux overlap function pseudo code**

```
for brdf1 = each in brdf_list1
  for brdf2 = each in brdf_list2
    if brdf1 != brdf2
      for each implicit
        for each x coordinate on 2D lattice
          for each y coordinate on 2D lattice
            for each lambda on wavelength
              flux1 = render(x, y, lambda, brdf1, implicit)
              flux2 = render(x, y, lambda, brdf2, implicit)
              error = FO(flux1, flux2)
              error_sum += error
            end for
          end for
        end for
      end for
    end if
  end for
end for
```

**Figure 6, flux overlap test pseudo code**

Analyzing the results, (see Figure 7) we see that Ward in combination with Ashikhmin is paramount, further candidates include Ashikhmin in combination with Strauss and Ashikhmin in combination with He-Torrance. Combining Phong with the Lafortune BRDF (see Figure 8) is 28040.33 times worse than Ward in combination with Ashikhmin and (see Figure 9). Seven combinations completely overlap and receive a result of zero. Note that we did not compare two of the same BRDF models since results would always be zero (perfect overlap). Also, note that there is not a one-on-one relation between complex physically plausible, anisotropic BRDF models and simple isotropic models, Strauss (ranked 5[th]) for example in combination with Ashikhmin receives a relative good result.

The results clearly indicate that there is a significant relationship between BRDF characteristics and total flux overlap:

o    Highest error in flux overlap implies two different anisotropic physically plausible BRDF models with Fresnel behavior.
o    High to medium error in flux overlap implies two different BRDF models where at least one BRDF model is physically plausible, anisotropic and includes Fresnel behavior.
o    Low error in flux overlap implies two different isotropic BRDF models each without physically plausible characteristics and without Fresnel behavior.

Please refer to appendix A for a complete analysis results overview.

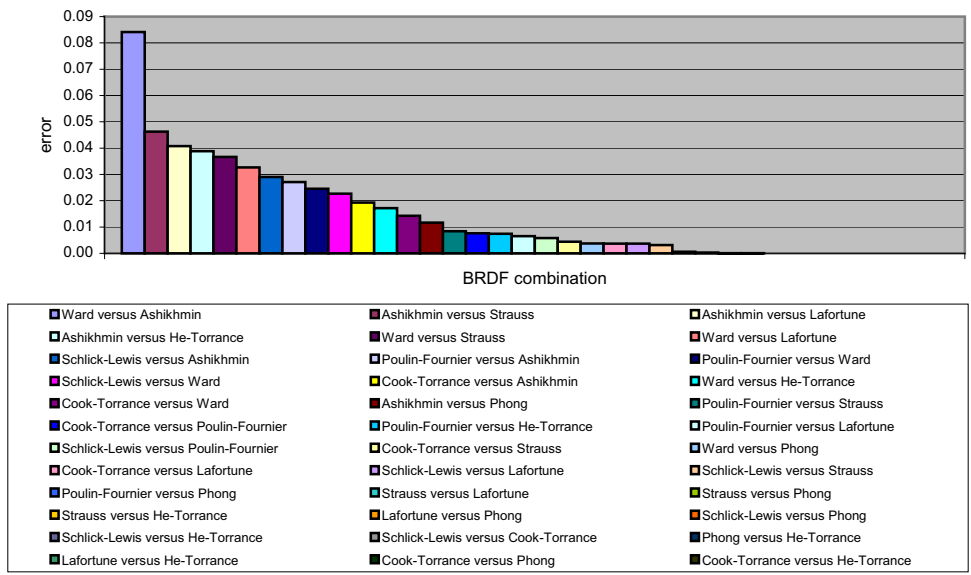

**Figure 7, BRDF flux overlap test results**

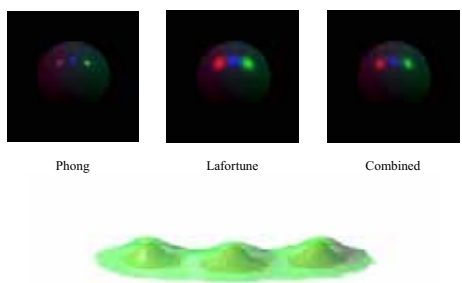

Phong     Lafortune     Combined

Combined three 3D specular lobes, green is Lafortune, red is Phong

**Figure 8, Worst case: Phong in combination with Lafortune**

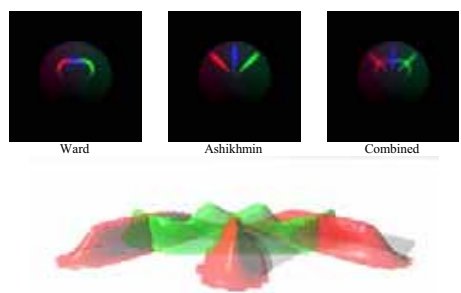

Ward     Ashikhmin     Combined

Combined three 3D specular lobes, green is Ashikhmin, red is Ward

**Figure 9, Best case: Ward in combination with Ashikhmin**

## Genetics study experiment

*Magnetic resonance imaging* (MRI) is an imaging technique used primarily in medical settings to produce high quality images of the inside of the human body. Magnetic resonance imaging is based on the absorption and emission of energy in the radio frequency range of the electromagnetic spectrum. MRI is based on the principles of nuclear magnetic resonance (NMR), a spectroscopic technique to obtain microscopic, chemical and physical information about molecules. MRI started out as a tomographic imaging technique; it produced an image of the NMR signal in a thin slice through the human body. MRI has advanced beyond a tomographic imaging technique to a volume imaging technique. The volume is composed of several volume elements or *voxels*. The volume of a voxel represents approximately between one and three $mm^3$ of tissue. The magnetic resonance image is composed of several picture elements called pixels. The intensity of a pixel is proportional to the NMR signal intensity of the contents of the corresponding volume element or voxel of the object being imaged.

At the Structural Neuroimaging Section of the University Medical Center Utrecht (UMCU), studies are performed into the genetic influences on variation in human brain morphology in health and disease. In a recent study, Hulshoff Pol et al [54] investigated the relationship between genetic factors and human brain morphology in the context of schizophrenia. The focus of the research is to determine if local changes in brain volume are related to genetic factors. Since the significant portions of the morphology of the genetic differences are small in a geometric sense and overlap, we chose to use the results of this study to experiment with a variation in reflection type for the different objects.

### Genetics study research

To assess the influence of genetic, common environmental and unique environmental factors on focal (sub) cortical gray matter and focal white matter volume, MRI images of 112 pairs of twins and 34 of their siblings were acquired. See [63] for a related and recent study in this field and [64] for the behavioral aspects.

### Data acquisition

Magnetic resonance images were acquired on a Philips NT scanner operating at 1.5 T in all subjects. T1-weighted 3D fast field echo (3D-FFE) scans with 160-180 1.2 mm contiguous coronal slices, and T2-weighted dual echo turbo spin echo (DE-TSE) scans with 120 1.6 mm contiguous coronal slices, of the whole head were used for quantitative measurements. In addition, T2-weighted DE-TSE scans with 17 axial 5 mm slices and 1.2 mm gap were used for clinical neuro-diagnostic evaluation.

### Data analysis

All images were coded to ensure blindness for subject identification and diagnosis. Scans were reoriented and aligned with the Talairach frame, and corrected for inhomogeneities in the magnetic field [59]. Binary masks of gray and white matter were made based on histogram analyses and series of mathematical morphological operators to connect all voxels of interest within the cranium [60]. Gray and white matter density maps were made and analyzed using voxel based morphometry. The gray and white matter density maps represent the local concentration of gray and white matter (between 0 and 1) per voxel throughout each of the 258 MRI brain scans, after transformation onto a standardized brain model to globally align anatomical regions while minimally affecting local volume changes. Refer to [65] for more information concerning the analysis stage.

### The cortex dataset

The standardized brain to which these images were transformed was selected earlier among 200 brain images of healthy subjects between 16-70 years of age. To select the standardized brain, all 200 brain images were registered to the Montreal standard brain [62] and averaged yielding one average brain image. The mean square error (MSE) on the normalized intensity values was computed between each of the brain images and the average brain image. The standardized brain was the brain image with the smallest MSE. Linear and non-linear transformations were then applied to the gray matter density maps to remove global differences in the size and shape of individual brains. Only the cortex of the standardized brain was used in our renderings. Since the standardized brain dataset contained highly varying tangent and normal orientations, it was resampled using a cubic B-spline approximation to four times its original size.

### The density map dataset

The binary gray matter masks were resampled to a voxel size of 2x2x2.4 mm$^3$, blurred using an isotropic Gaussian kernel (full width at half maximum of 8 mm) to generate gray matter "density maps". The density maps represent the local concentration of gray matter (between 0 and 1) per voxel. Each of the MRI images was transformed into a standardized coordinate system in a two-stage process using the ANIMAL algorithm [58]. In the first step, a linear transformation was found by minimizing a mutual information joint entropy objective function computed on the gray level images [61]. A nonlinear transformation was computed in the second step by maximizing the correlation of the subject's image with that of a standardized brain. The nonlinear transformation is run up to a scale that aligns global anatomical regions while minimally affecting local volume changes.

The density map was separated into four significant areas with varying iso surface values (see Figure 10) related to genetic influence in gray matter density map significance ranging from low importance (0.70) to high (0.85).

| Iso | 0.70 | 0.75 | 0.80 | 0.85 |
|---|---|---|---|---|
| Name | Density map 0 | Density map 1 | Density map 2 | Density map 3 |
| Image | | | | |
| Opacity | 0.3 | 0.4 | 0.5 | 0.6 |
| Color | Blue | Green | Yellow | Red |

**Figure 10, Density map iso threshold separation**

### The CThead dataset

The CThead [69] is not part of the genetics study; it is only used to visually place the cortex and the density maps in the correct geometric context. The dataset was used with courtesy from the SoftLab Software Systems Laboratory, University of North Carolina. The CT cadaver head data is a 113-slice MRI data set of a CT study of a cadaver head. Only the iso surface of the skull (0.5) was used in our renderings.

### Rendering settings

A scene was constructed in which all objects (see Figure 11) were placed at appropriate positions and all objects were rotated by 45 degrees (front-side view). Voxel sampling was set to 0.05, i.e. for every voxel 20 samples were taken. Normal estimation was based on a cubic B-spline algorithm for the cortex and central differences for other objects – rapid changing tangent orientations can influence light-matter interaction therefore normal estimation is essential, see [29] for an excellent overview of normal estimation methods. Four omni lights were added all behind the camera in each corner. The CThead was rendered using Direct Surface Rendering (DSR) with an iso-surface value of 0.5 and opacity of 0.97 and assigned a specific BRDF. Both density maps and cortex were rendered using Direct Volume Rendering (DVR) with ramped transfer functions for both opacity and color and each was assigned a specific BRDF, the opacity of the cortex ranged from 0.001 to 0.2. A Lambertian diffuse constant of 0.2 was added to each object for geometric attenuation. The datasets were rendered against a random colored background to avoid the Gelb effect. Blur was added to the background to separate the dataset from the background in depth. The background was rendered with a standard Phong BRDF model where only ambient (Ka 0.2) and diffuse (Kd 0.3) illumination was used. Because the same background was used for each rendering, it provides no methodical information about the illumination. Images were rendered into an image of size 1024x1024 and resampled from floating point to RGB color space. Each pixel was super-sampled eight times using jittered sampling. Pre-processing (interpolation) of datasets was done on a HP Unix k200 server. All images were rendered on a 666Mhz Pentium-3 computer running Linux taking approximately 20 hours per image.

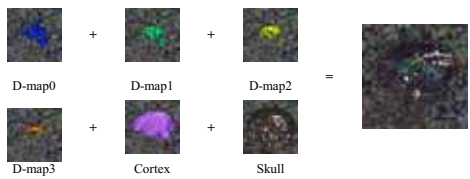

**Figure 11, Object setup**

## Results

Based on and our selection criteria and total flux overlap results we selected the three best and the three worst candidates. The first six samples are based on all permutation of the three best selected (combination A, see Table 2). The second six samples are based on the permutations of the three worst candidates (combination B, see Table 3). The next six samples are not differentiated, i.e. one of the six BRDF models was used for all objects (combination C, see Table 4). As noted previously, some combinations of a weaker candidate with a strong candidate perform well in the total flux overlap test, thus additional renderings were made with combinations of the latter (combination D, see Table 5). Additional high-resolution renderings were made where the camera zoomed into the density-map region to examine the extent of visibility of anatomical features.

| Rendering | CThead BRDF | Cortex BRDF | Density maps BRDF |
|---|---|---|---|
| 1 | Ward | Ashikhmin | He-Torrance |
| 2 | Ward | He-Torrance | Ashikhmin |
| 3 | Ashikhmin | Ward | He-Torrance |
| 4 | Ashikhmin | He-Torrance | Ward |
| 5 | He-Torrance | Ward | Ashikhmin |
| 6 | He-Torrance | Ashikhmin | Ward |

**Table 2, BRDF combination (A)**

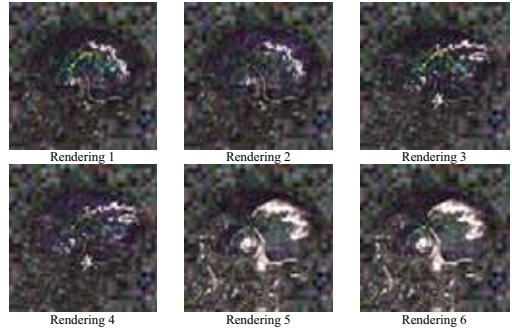

**Figure 12, combination (A) renderings**

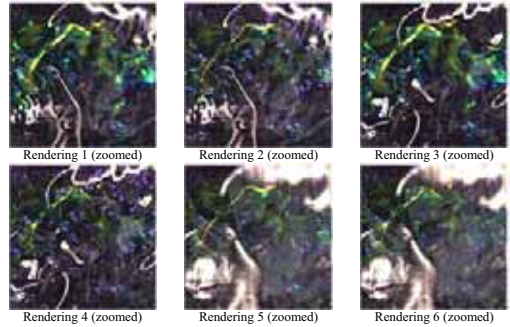

**Figure 13, combination (A) renderings, zoomed**

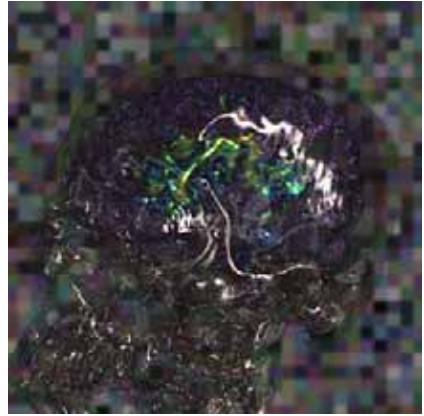

**Figure 14, enlargement of rendering 1**

| Rendering | CThead BRDF | Cortex BRDF | Density maps BRDF |
|---|---|---|---|
| 7 | Cook-Torrance | Schlick-Lewis | Phong |
| 8 | Cook-Torrance | Phong | Schlick-Lewis |
| 9 | Schlick-Lewis | Cook-Torrance | Phong |
| 10 | Schlick-Lewis | Phong | Cook-Torrance |
| 11 | Phong | Schlick-Lewis | Cook-Torrance |
| 12 | Phong | Cook-Torrance | Schlick-Lewis |

**Table 3, BRDF combination (B)**

| Rendering | CThead BRDF | Cortex BRDF | Density maps BRDF |
|---|---|---|---|
| 13 | Ward | Ward | Ward |
| 14 | Ashikhmin | Ashikhmin | Ashikhmin |
| 15 | He-Torrance | He-Torrance | He-Torrance |
| 16 | Cook-Torrance | Cook-Torrance | Cook-Torrance |
| 17 | Schlick-Lewis | Schlick-Lewis | Schlick-Lewis |
| 18 | Phong | Phong | Phong |

**Table 4, BRDF combination (C)**

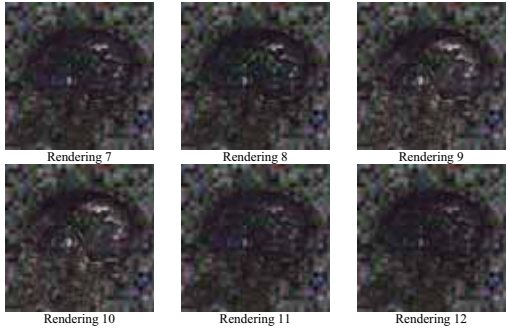

**Figure 15, combination (B) renderings**

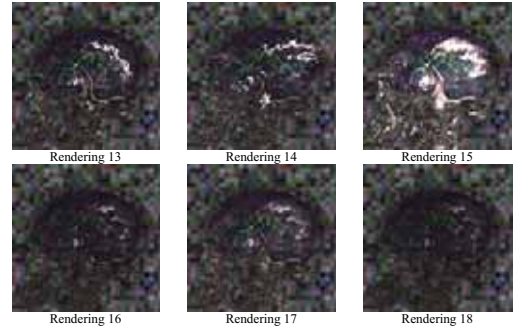

**Figure 18, combination (C) renderings**

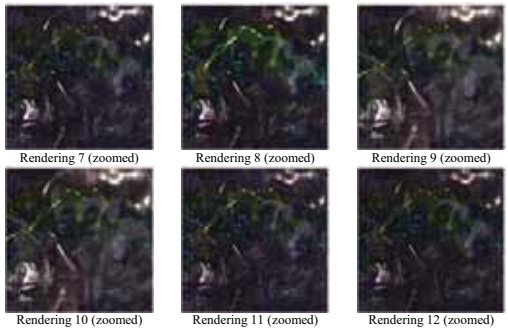

**Figure 16, combination (B) renderings, zoomed**

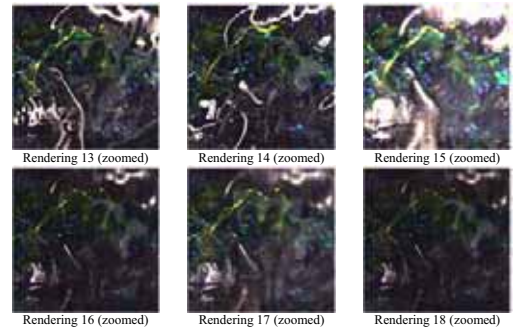

**Figure 19, combination (C) renderings, zoomed**

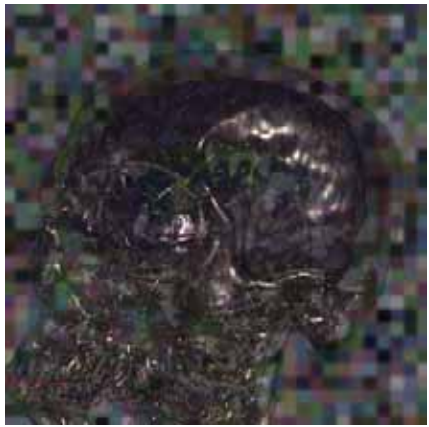

**Figure 17, enlargement of rendering 9**

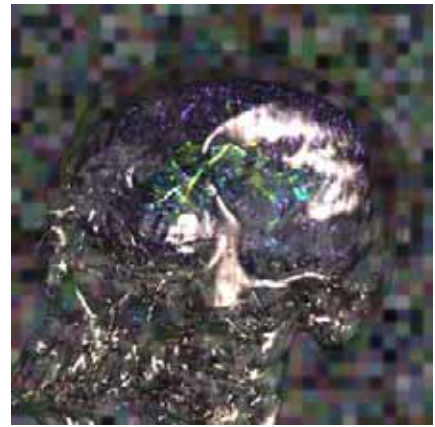

**Figure 20, enlargement of rendering 15**

| Rendering | CThead BRDF | Cortex BRDF | Density maps BRDF |
|---|---|---|---|
| 19 | Ward | Schlick-Lewis | Phong |
| 20 | Ward | Phong | Schlick-Lewis |
| 21 | Schlick-Lewis | Ward | Phong |
| 22 | Schlick-Lewis | Phong | Ward |
| 23 | Phong | Ward | Schlick-Lewis |
| 24 | Phong | Schlick-Lewis | Ward |

**Table 5, BRDF combination (D)**

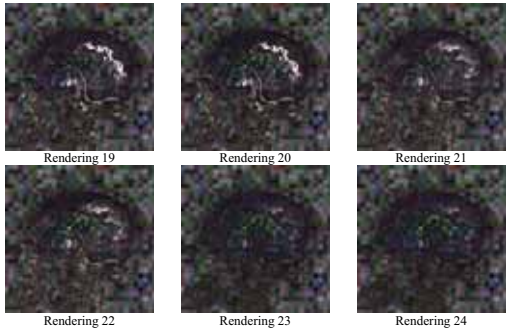

Rendering 19 — Rendering 20 — Rendering 21
Rendering 22 — Rendering 23 — Rendering 24

**Figure 21, combination (D) renderings**

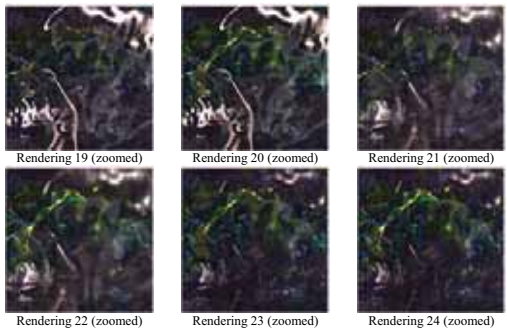

Rendering 19 (zoomed) — Rendering 20 (zoomed) — Rendering 21 (zoomed)
Rendering 22 (zoomed) — Rendering 23 (zoomed) — Rendering 24 (zoomed)

**Figure 22, combination (D) renderings, zoomed**

## Discussion

We set out to investigate if combining different BRDF models would enhance recognition and localization of structures within MRI data. We defined four criteria and selected nine BRDF models. The selected models were tested in combination with each other to analyze which combination yielded the best results in terms of specular flux overlap. The combination of models that ranked highest in the latter test was rendered along with the combination of models that ranked lowest. The results indicate that if a single isotropic, not physically plausible BRDF without Fresnel behavior is used for all objects in a scene, anatomical features are hard to recognize (see Figure 18 and Figure 19, rendering 16 through 18). Furthermore, differentiating simple isotropic BRDF models, that are not physically plausible and do not include Fresnel behavior give slightly better results, yet much detail is hidden and objects are still hard to recognize (see Figure 15 and Figure 16). Visibility of objects dramatically increases if a physically plausible anisotropic BRDF model is introduced that includes Fresnel behavior, even if the other objects in the scene are simple isotropic BRDF models (Figure 21 and Figure 22). Another important observation is that using a single physically plausible anisotropic BRDF model with Fresnel behavior for all objects is not paramount; although results are enhanced, some anatomical detail is still hidden (see Figure 18 and Figure 19, rendering 13 through 15). The enhanced results are a direct consequence of the physically plausible nature and Fresnel behavior of the BRDF models, not of the anisotropic behavior since similar anisotropic models would still yield a low total flux overlap. For improved results the BRDF models should not only be physically plausible, anisotropic and include Fresnel behavior, they should also be differentiated (see Figure 12 and Figure 13) and differentiation should depend on total flux overlap for the combination of BRDF models. Another important factor in the context of visibility is the order of BRDF models in relation to objects; if simple BRDF models are applied to outer hull objects and inner objects contain a complex BRDF model results will likely be unsatisfactory while reversing BRDF models will likely yield better results (notice the differences in Figure 21 and Figure 22).

Beside *esthetics,* there are significant differences between renderings. Notice that in some renderings all objects are visible while in other renderings hardly any object is visible. Observe that in Figure 12 and Figure 13, all the renderings portray all objects, i.e. the skull, the cortex and the density maps are clearly visible, suggesting that differentiating reflection models is worthwhile. Which of the latter six renderings is the most *intuitive* is a question of personal taste. Renderings in Figure 18 and Figure 19 shows that using an anisotropic model for all objects (renderings 13 through 15) is better than using an isotropic model for all objects (renderings 16 through 18), yet it still does not show *all* objects; in renderings 16 through 18, nothing is particularity visible, object reflections overlap and light-matter interaction is significantly worse.

We measured the rendering time for each visualization, the differences in computational time for simple and complex BRDF models is only small percentage wise: using only a computationally inexpensive BRDF model (Schlick-Lewis, timed as 40m:20s in low resolution and 19h:23m:12s in high resolution) for all objects is only three percent faster than a computationally expensive BRDF model (He-Torrance, timed as 41m:23s in low resolution and 19h:59m:33s in high resolution). The evaluation of the BRDF equation is reasonably trivial in computation time compared to all other stages in our volume-rendering pipeline.

Refer to appendix A, Table 7 for individual BRDF model rendering times.

## Conclusion

The aim of our research is to investigate if reflection models play an essential role in the visibility and localization of certain objects within a collection of more objects. We investigated different aspects of individual BRDF models including anisotropic behavior, Fresnel behavior and physical plausibility and empirically examined BRDF combinations by measuring total flux overlap.

We conclude:

1. If all the BRDF models are isotropic, not physically plausible and repeatedly used for all objects in the scene, internal structures and object locations are hard to detect.

2. If one of the BRDF models is anisotropic, physically plausible and includes Fresnel behavior, internal structures and object locations are significantly easier to detect.

3. To achieve dominant results, BRDF models should not only be anisotropic, physically plausible and contain Fresnel behavior but should also be differentiated for each object, and differentiation should be based on total flux overlap error.

The results of our experiment indicate that using only simple isotropic Phong or Phong-alike reflection for all objects is possibly the worst combination to use; yet, most volume renders in use today employ just that. Computational time does not seem to be major factor in evaluating complex differentiating BRDF models, since it is only a small part of the total volume visualization pipeline.

The question is *when* BRDF differentiation becomes important. Differentiation is valuable once data is complex, however a definition of complex is hard to characterize. In summary, reflection models are important for the ability of humans to visually separate objects and differentiating reflection models is worthwhile if anatomical feature recognition is important.

## Future work

We have only taken the first steps in investigating the relationship between reflection models and increase in object visibility. Further research should include studying the relationship between the more general Bi-directional Scattering Distribution Function (BSDF) and Bi-directional Subsurface Scattering Distribution Function (BSSRDF) and investigate if other BRDF models, especially measured BRDF models, give improved results. One could also hypothesize if building an *adaptive BRDF*, i.e. a BRDF model that adapts itself to other active BRDF models in a scene given a certain total flux overlap threshold, would avoid flux overlap to occur in the first place. Further research into the flux overlap definition is also warranted; currently no score is given to distance or geometric shape of the reflection lobe while humans may possibly be more convinced by differentiating form and distances between lobes. More importantly, one can question which attribute is the most significant for humans to decide what and where an object is located among other objects. Is it spectral information (color) or reflection information, and in what measure? We are looking into the possibility of setting up a controlled measurement environment in which neurological expert subjects should determine which rendering is the most '*intuitive*', i.e. which combination of BRDF models is paramount in the context of localizing specific anatomical features. A possibly interesting direction is to devise a test in which subjects are asked to specify volume size and location of specific structures within complex anatomy from a set of different rendered models. Parameters for each BRDF model also play an important role, although in our research we used static parameters one could envision that altering (either statically or dynamically) parameters would significantly change the outcome. Some work is currently in progress at the Institute for Computer, Communication and Media Technology; three students are currently investigating the relationship between organic materials and light-interaction, adapting photon mapping to MRI data and studying the relationships between NPR rendering schemes and visibility. The results of these studies should give further insight into the relation between light-material interaction and MRI data interpretation.

## Acknowledgements

Thanks to Michael Ashikhmin, University of Utah, for his tips on getting the Ashikhmin BRDF to work properly. Thanks to Fern Y. Hunt, National Institute of Standards and Technology, Mathematical and Computational Sciences Division for introducing us to the wonders of the Non-conventional Exploitation Factors Data System (NEFDS) and according generated BRDF data. Many thanks to Jason Ang, University of Waterloo for his patience in trying to explain what was going wrong in the anisotropic BRDF computations. Also thanks to Philip Dutré, Computer Graphics

Group, Department of Computer Science, K.U. Leuven for his tip on energy conservation in the Phong lightning equation. Thanks to Hugo Schnack, University Utrecht, for re-sampling the cortex dataset. Big thanks to my advisor René Mandl, University Utrecht, for his stimulation, brainstorm sessions and helping me get on the right track – and more importantly keeping me there. Finally, many thanks to Hilleke Hulshoff Pol, University Utrecht, for giving me the opportunity to perform this study in the first place and for allowing me to use the genetics study data as a basis for testing my ideas.

## Appendix A

Table 6 contains all measures from the flux overlap test. Table 7 contains timings of each individual BRDF for an image size of 512x512 and only one object (the CThead).

| Rank | Error | BRDF combination |
|---|---|---|
| 1 | 0.084121 | Ward versus Ashikhmin |
| 2 | 0.046296 | Ashikhmin versus Strauss |
| 3 | 0.040771 | Ashikhmin versus Lafortune |
| 4 | 0.038811 | Ashikhmin versus He-Torrance |
| 5 | 0.036682 | Ward versus Strauss |
| 6 | 0.032664 | Ward versus Lafortune |
| 7 | 0.029034 | Schlick-Lewis versus Ashikhmin |
| 8 | 0.027088 | Poulin-Fournier versus Ashikhmin |
| 9 | 0.024580 | Poulin-Fournier versus Ward |
| 10 | 0.022679 | Schlick-Lewis versus Ward |
| 11 | 0.019312 | Cook-Torrance versus Ashikhmin |
| 12 | 0.017198 | Ward versus He-Torrance |
| 13 | 0.014297 | Cook-Torrance versus Ward |
| 14 | 0.011677 | Ashikhmin versus Phong |
| 15 | 0.008398 | Poulin-Fournier versus Strauss |
| 16 | 0.007648 | Cook-Torrance versus Poulin-Fournier |
| 17 | 0.007432 | Poulin-Fournier versus He-Torrance |
| 18 | 0.006559 | Poulin-Fournier versus Lafortune |
| 19 | 0.005779 | Schlick-Lewis versus Poulin-Fournier |
| 20 | 0.004412 | Cook-Torrance versus Strauss |
| 21 | 0.003778 | Ward versus Phong |
| 22 | 0.003719 | Cook-Torrance versus Lafortune |
| 23 | 0.003660 | Schlick-Lewis versus Lafortune |
| 24 | 0.003131 | Schlick-Lewis versus Strauss |
| 25 | 0.000580 | Poulin-Fournier versus Phong |
| 26 | 0.000275 | Strauss versus Lafortune |
| 27 | 0.000083 | Strauss versus Phong |
| 28 | 0.000066 | Strauss versus He-Torrance |
| 29 | 0.000003 | Lafortune versus Phong |
| 30 | 0.000000 | Schlick-Lewis versus Phong |
| 30 | 0.000000 | Schlick-Lewis versus He-Torrance |
| 30 | 0.000000 | Schlick-Lewis versus Cook-Torrance |
| 30 | 0.000000 | Phong versus He-Torrance |
| 30 | 0.000000 | Lafortune versus He-Torrance |
| 30 | 0.000000 | Cook-Torrance versus Phong |
| 30 | 0.000000 | Cook-Torrance versus He-Torrance |

**Table 6, total flux overlap error results**

| BRDF model | Rendering time |
|---|---|
| Ward | h00:m57:s42 |
| Ashikhmin | h00:m55:s54 |
| He-Torrance | h00:m56:s12 |
| Cook-Torrance | h00:m56:s12 |
| Schlick-Lewis | h00:m55:s01 |
| Phong | h00:m55:s12 |
| Strauss | h00:m55:s25 |
| Poulin-Fournier | h00:m54:s42 |
| Lafortune | h00:m56:s07 |

**Table 7, individual BRDF timings**

## Appendix B

Used parameters for each BRDF in the context of this research were:

**[1] Cook-Torrance BRDF**

Surface roughness was set to 0.08, index of refraction (real part) 1.6, index of refraction (imaginary part) 0.2 and specular reflectivity 0.8.

**[2] Poulin-Fournier BRDF**

Distance between cylinders was set to 2.0, height of floor between cylinders 4.0, shininess of surface 100.0 and specular reflectivity of surface 0.8.

**[3] Ward BRDF**

Roughness in X direction was set to 0.05, roughness in Y direction 0.3 and specular reflectivity 0.05.

**[4] Ashikhmin BRDF**

Specular u exponent was set to 10 and specular v exponent 1000.

**[5] Strauss BRDF**

Smoothness of surface was set to 0.75, metalness of surface 0.5 and specular reflectivity 0.5. The original Strauss magic constants were used.

**[6] Lafortune BRDF**

Cx parameter was set to -1.0, Cy parameter -1.0, Cz parameter 0.95 and cosine exponent 20.0. One lobe was used.

**[7] Phong BRDF**

Cosine exponent (n) was set to 10 and specular constant 0.8.

**[8] He-Torrance BRDF**

Tau/lambda in He's model was set to 10.0, sigma/lambda in He's model 1.0, index of refraction (real part) 1.6, index of refraction (imaginary part) -0.2 and specular reflectivity 0.8.

**[9] Schlick-Lewis specular BRDF**

Cosine exponent (n) was set to 10 and specular constant 0.8.

## References

[1] Reza Kasrai, Frederick A. A. Kingdom, and Terry M. Peters: The Perception of Transparency in Medical Images, pages 726-733.

[2] Herke Jan Noordmans, Hans T. M. van der Voort, and Max A. Viergever: Modeling Spectral Changes to Visualize Embedded Volume Structures for Medical Image Data, pages 706-715.

[3] Herke Jan Noordmans, Hans T.M. van der Voort, and Arnold W.M. Smeulders: Spectral Volume Rendering. IEEE transactions on visualization and computer graphics, vol. 6, no. 3, july-september 2000.

[4] Helwig Hauser, Lukas Mroz, Gian Italo Bischi, and M. Eduard Gröller: Two-Level Volume Rendering. IEEE transactions on visualization and computer graphics, vol. 7, no. 3, july-september 2001.

[5] Markus Hadwiger, Helwig Hauser: First Steps in Hardware Two-Level Volume Rendering, VrVis 2002.

[6] Jianlong Zhou, Manfred Hinz, and Klaus D. Tönnies: Hybrid Focal Region-Based Volume Rendering of Medical Data.

[7] Pierre Poulin and Alain Fournier: A Model for Anisotropic Reflection, Computer Graphics, volume 24, pages 273-282 1990.

[8] J. T. Kajiya: The rendering equation, Computer Graphics (Siggraph '86 Proceedings), volume 20, pages 143-150, (Aug. 1986). The Eurographics Association and Blackwell Publishers 1999.

[9] Christophe Schlick: A Customizable Reflectance Model for Everyday Rendering, Fourth Eurographics Workshop on Rendering, Series EG 93 RW, pages 73-84, 1993.

[10] J. F. Blinn: Models of light reflection for computer synthesized pictures, Computer Graphics Siggraph, 11, 1977.

[11] Michael Ashikhmin Peter Shirley: An Anisotropic Phong Light Reflection Model, 2000.

[12] L. Neumann and A. Neumann: Photo simulation: Interreflection with Arbitrary Reflectance Models and Illuminations, Computer Graphics Forum, volume 8, pages 21–34, 1989.

[13] Stephen R. Marschner and Richard J. Lobb: An Evaluation of Reconstruction Filters for Volume Rendering, Proceedings of Visualization '94, R. Daniel Bergeron and Arie E. Kaufman, pages 100-107, 1994.

[14] Arie Kadosh and Daniel Cohen-Or and Omer Shibolet and Roni Yagel: Rendering Discrete Surfaces from Close-up Views.

[15] Foley, J. D., van Dam, A., Feiner, S. K., and Hughes, J. F.: Computer Graphics - principles and practice, 2nd ed. Reading, Massachusetts: Addison Wesley, 1990.

[16] James F. Blinn: Models of light reflection for computer synthesized pictures, Proceedings of the 4th annual conference on Computer graphics and interactive techniques, pages 192-198, July 20-22, 1977.

[17] Hanspeter Pfister and Jan Hardenbergh and Jim Knittel and Hugh Lauer and Larry Seiler: The VolumePro Real-Time Ray-casting System, Computer Graphics Proceedings Addison Wesley Longman, Addison Wesley Longman, pages 251-260, 1999.

[18] Philippe Lacroute and Marc Levoy: Fast Volume Rendering Using a Shear-Warp Factorization of the Viewing Transformation, Computer Graphics volume 18, pages 451-458, 1994.

[19] Nicodemus, F. E., Richmon, J. C., Hsia, J. J., Ginsberg, I. W., and Limperis, T.: Geometric considerations and nomenclature for reflectance. NBS Monograph 160, National Bureau of Standards, Washington, DC, 1977.

[20] Stephen H. Westin, James R. Arvo, Kenneth E. Torrance: Predicting reflectance functions from complex surfaces, ACM Siggraph Computer Graphics, volume 26 no 2, pages 255-264, July 1992.

[21] Donald P. Greenberg: A framework for realistic image synthesis, Communications of the ACM, volume 42 no 8, pages 44-53, Aug. 1999.

[22] Jan Kautz, Hans-Peter Seidel: Towards interactive bump mapping with anisotropic shift-variant BRDFs, Proceedings 2000 Siggraph/Eurographics workshop on on Graphics hardware, pages 51-58, August 21-22, 2000.

[23] M. Goss: An Adjustable Gradient Filter for Volume Visualization Image Enhancement, Proceedings Graphics Interface, pages 67-74, 1994.

[24] T. Moller, R. Machiraju, K. Mueller, and R. Yagel: Evaluation and Design of Filters Using a Taylor Series Expansion, IEEE Trans. Visualization and Computer Graphics, volume 3, no. 2, pages 184-199.

[25] T. Moller, K. Mueller, Y. Kurzion, R. Machiraju, and R. Yagel: Design of Accurate and Smooth Filters for Function and Derivative Reconstruction, Proc. 1998 IEEE Symposium, Volume Visualization, pages 143-151, 1998.

[26] S. Marschner and R. Lobb: An Evaluation of Reconstruction Filters for Volume Rendering, Proceedings IEEE Visualization pages 100-107, 1994.

[27] László Neumann, Balázs Csébfalvi, A. König and E. Gröller: Gradient Estimation in Volume Data using 4{D} Linear Regression, Computer Graphics Forum (Eurographics 2000), volume 19(3), pages 351-358, 2000.

[28] M. Bentum, B. Lichtenbelt, and T. Malzbender: Frequency Analysis of Gradient Estimators in Volume Rendering, IEEE Trans. Visualization and Computer Graphics, vol. 2, no. 3, Sept. 1996.

[29] T. Möller, R. Machiraju, K. Mueller, and R. Yagel: A comparison of normal estimation schemes, IEEE Visualization, pages 19-26, Phoenix, October 1997.

[30] T. Möller, R. Machiraju, K. Mueller, and Roni Yagel: Classification and local error estimation of interpolation and derivative filters for volume rendering, 1996 Symposium on Volume Visualization, pages 71-78, San Francisco, October 1996.

[31] T. Möller, R. Machiraju, K. Mueller, and Roni Yagel: Evaluation and design of filters using a Taylor series expansion, IEEE Transactions on Visualization and Computer Graphics vol. 3, no. 2, pages 184-199, 1997.

[32] Barthold Lichtenbelt, Randy Crane, and Shaz Naqvi: Introduction to Volume Rendering, chapter 4. Prentice-Hall, New Jersey, 1998.

[33] Wolgang Krueger: The application of transport theory to visualization of 3-D scalar data fields, Computers in Physics, pages 397–406, July-August 1991.

[34] Gordon Kindlmann and David Weinstein: Hue-Balls and Lit-Tensors for Direct Volume Rendering of Diffusion Tensor Fields. In Proceedings Visualization 1999, pages 183–189, October 1999.

[35] David Rodgman and Min Chen: Refraction in Discrete Ray Tracing. In Proceedings Volume Graphics 2001, pages 3–17, 403, June 2001.

[36] Richard S. Hunter and Richard W. Harold: The measurement of appearance.

[37] Roland W. Fleming, Ron O. Dror, and Edward H. Adelson: How do Humans Determine Reflectance Properties under Unknown Illumination, Massachusetts Institute of Technology.

[38] Ron O. Dror: Surface Reflectance Recognition and Real-World Illumination Statistics.

[39] Robert Lewis: Making shaders more physically plausible. In Michael F. Cohen, Claude Puech, and Francois Sillion, editors, Fourth Eurographics Workshop on Rendering, pages 47-62, Eurographics, June 1993, held in Paris, France, 14-16 June 1993.

[40] Andrew S. Glassner: Principles of Digital Image Synthesis, volume 2. Morgan Kaufmann, San Francisco, 1995.

[41] Gregory J. Ward: Measuring and Modeling Anisotropic Reflection. In ACM Siggraph 1992 Conference Proceedings, volume 26, pages 265–272, July 1992.

[42] Eric Lafortune, Sing-Choong Foo, Kenneth E. Torrance, and Donald P.Greenberg: Non-linear Approximation of Reflectance Functions. In ACM Siggraph '97 Conference Proceedings, pages 117–126, August 1997.

[43] Kenneth E. Torrance and E.M. Sparrow: Theory for off Specular Reflection from Roughened Surfaces. Journal of the Optical Society of America, 57, September 1967.

[44] Robert L. Cook and Kenneth E. Torrance: A Reflectance Model for Computer Graphics. In ACM Siggraph '81 Conference Proceedings, volume 15, pages 307–316, August 1981.

[45] Xiao D. He, Kenneth E. Torrance, Francois X. Sillion and Donald P. Greenberg: A comprehensive Physical Model for Light Reflection. In ACM Siggraph '91 Conference Proceedings, volume 25, pages 175–186, July 1991.

[46] Pierre Poulin and Alain Fournier: A Model for Anisotropic Reflection, In ACM Siggraph '90 Conference Proceedings, volume 24, pages 273–282, 1990.

[47] Peter Shirley: Realistic Ray Tracing.

[48] Aidong Lu and Christopher J. Morris and David S. Ebert and Penny Rheingans and Charles Hansen: Non-Photorealistic Volume Rendering Using Stippling Techniques.

[49] David Ebert and Penny Rheingans: Volume Illustration: Non-Photorealistic Rendering of Volume Models, Proceedings Visualization 2000, T. Ertl and B. Hamann and A. Varshney, pages 195-202, 2000.

[50] Treavett, S. M. F., Chen, M., Satherley, R., Jones, M. W: Volumes of Expression: Artistic Modelling and Rendering of Volume Datasets", Proc. of Computer Graphics International, Hong Kong, July 2001.

[51] Gordon Kindlmann and James W. Durkin: Semi-Automatic Generation of Transfer Functions for Direct Volume Rendering, IEEE Symposium on Volume Visualization, pages 79-86, 1998.

[52] Joe Kniss and Patrick McCormick and Allen McPherson and James Ahrens and Jamie Painter and Alan Keahey and Charles Hansen: Interactive Texture-Based Volume Rendering for Large Data Sets, IEEE Computer Graphics and Applications, volume 21, number 4, pages 52-61, 2001.

[53] Christophe Schlick: An Inexpensive BRDF Model for Physically-Based Rendering, Computer Graphics Forum, volume 13, number 3, pages 233-246, 1994.

[54] Hulshoff Pol et al: Focal morphology of brain regions involved is genetically determined, in preparation.

[55] Bui Tong Phong: Illumination for Computer Generated Images, Communications of the ACM Volume 18 Issue 6, pages 311-317, June 1975.

[56] Paul S. Strauss: A Realistic Lighting Model for Computer Animators, IEEE Computer Graphics and Applications, pages 56-64, November 1990.

[57] Henrik Wann Jensen: Realistic Image Synthesis Using Photon Mapping.

[58] Collins DL, Holmes CJ, Peters TM & Evans AC: Automatic 3-D Model-Based Neuroanatomical Segmentation, Human Brain mapping, volume 4, pages 190-208, 1996.

[59] Sled JG, Zijdenbos AP & Evans AC.: A nonparametric method for automatic correction of intensity nonuniformity in MRI data. IEEE Transactions on Medical Imaging, volume 17, pages 87-97, 1998.

[60] Schnack HG, Hulshoff Pol HE, Baaré WFC, Staal WG, Viergever MA, Kahn RS: Automated separation of gray and white matter from MR images of the human brain, Neuroimage, volume 13, pages 230-237, 2001.

[61] Maes HH, Neale MC, Eaves IJ: Genetic and environmental factors in relative body weight and human adiposity, Behavioral Genetics volume 27, pages 325-351, 1997.

[62] Collins, D. L., Neelin, P., Peters, T. M. and Evans, A. C.: Automatic 3D inter-subject registration of MR volumetric data in standardized talairach space, JCAT, volume 18, pages 192-205, 1994.

[63] Hilleke E. Hulshoff Pol, Hugo G. Schnack, René C.W. Mandl, Neeltje E. M. van Haren, Hilde Koning, D. Louis Collins, Alan C. Evans, René S. Kahn: Focal gray matter density changes in schizophrenia, Arch Gen Psychiatry, volume 58, pages 1118-1125, 2001.

[64] Posthuma D, de Geus EJC, Neale MC, Hulshoff Pol HE, Baaré WFC, Kahn RS, Boomsma DI: Multivariate genetic analysis of brain structure in an extended twin design, Behavior Genetics, volume 30, pages 311-319, 2000.

[65] Baare WFC, Hulshoff Pol HE, Boomsma DI, Posthuma D, de Geus EJC, Schnack HG, van Haren NEM, van Oel CJ, Kahn RS: Quantitative genetic modeling of variation in human brain morphology, Cerebral Cortex, volume 11, pages 816-824, 2001.

[66] Helmholtz, H. v: Treatise on physiological optics, New York, Optical Society of America, 1866.

[67] Gelb, A: Die "Farbenkonstanz" der Sehdinge. In W. A. von Bethe (Ed.), Handbuch der Normal und Pathologische Psychologie, pages 594-678, 1929.

[68] Wallach, H.: Brightness constancy and the nature of achromatic colors, Journal of Experimental Psychology, volume 38, pages 310-324, 1948.

[69] http://www.siggraph.org/education/materials/vol-viz/volume_visualization_data_sets.htm